# Ultrasonic force microscopy on poly(vinyl alcohol)/SrTiO3 nano-perovskites hybrid films

Salvatore Marino[a], Girish M. Joshi[b], Angelo Lusuardi[a], M. Teresa Cuberes[a]

[a]*Laboratory of Nanotechnology, University of Castilla-La Mancha, Plaza Manuel Meca 1, 13400 Almadén, Spain*
[b]*Polymer Nanocomposite Laboratory, Materials Physics Division, School of Advanced Sciences VIT University, Vellore 14, Tamil Nadu, India*

ABSTRACT: Atomic Force Microscopy (AFM) and Ultrasonic Force Microscopy (UFM) have been applied to the characterization of composite samples formed by $SrTiO_3$ (STO) nanoparticles (NPs) and polyvinyl alcohol (PVA). The morphological features of the STO NPs were much better resolved in UFM than in contact-mode AFM topography. For high STO concentrations the individual STO NPs formed nanoclusters, which gathered in microaggregates. The STO aggregates, covered by PVA, exhibited no AFM frictional contrast, but were clearly distinguished from the PVA matrix using UFM. Similar aggregation was observed for NPs in the composite samples than for NPs deposited on top of a flat silicon substrate from a milliQ water solution in the absence of polymer. In the hybrid films, most STO nanoparticles typically presented a lower UFM contrast than the PVA matrix, even though stiffer sample regions such as STO should give rise to a higher UFM contrast. STO NPs with intermediate contrast were characterized by an UFM halo of lower contrast at the PVA/STO interface. The results may be explained by considering that ultrasound is effectively damped on the nanometer scale at PVA/ STO interfaces. According to our data, the nanoscale ultrasonic response at the PVA/STO interface plays a fundamental role in the UFM image contrast.

KEYWORDS. Atomic Force Microscopy. Ultrasonic Force Microscopy. Poly(vinyl alcohol). Stroncium Titanate. Nanoparticles. Nanocomposites.

**1- Introduction**

Ultrasonic Force Microscopy (UFM) is a powerful technique to investigate the elastic and adhesive response of materials on the nanoscale [1,2]. The procedure is capable to provide material contrast in both soft and hard samples, bringing additional advantages when compared with other Scanning Probe Microscopy (SPM) approaches [3-6]. Here, UFM is applied to the characterization of composite samples formed by $SrTiO_3$ (STO) nanoparticles (NP) and poly(vinyl alcohol) (PVA). We intend to further explore the capability of the technique to provide subsurface information, to characterize the nanostructures formed by STO NP in the PVA hybrid films, and to gain inside into the mechanisms of ultrasound propagation on the nanoscale and the origin of the UFM contrast.

The insertion of ceramics nanoparticles into polymer matrix has led to the generation of novel hybrid materials with improved electrical and thermo-mechanical properties. In titanate-polymer composites, the titanates contribute with a high capacitance, and the polymers are typically easy to process. Hence, the composite films appear very attractive for the fabrication of integrated circuits [7,8]. Composites of STO with polymeric materials have been considered for microwave applications [9, 10]. The dielectric properties of PVA mixed with $PbTiO_3$ show promise for their application as supercapacitors and humidity sensors [11].

STO nanoparticles are being tested for the development of thin film transistors [8], batteries [12], photodiodes [13], and solar cells [14]. Also, they may exhibit photocatalytic activity [15-17]. The nanoparticle optical responses are highly dependent on their size and doping state [18,19]. The electronic properties of transition metal oxides interfaces are currently attracting a great deal of interest for device engineering [20]. Recently, a field effect transistor device has been implemented on an STO single crystal with a PVA gate insulator layer [21]. In bulk and pure form, STO remains paraelectric down to 0 K, even though chemical or isotopic substitution, or the application of stress, may easily disturb this state, resulting in ferroelectricity [22].

PVA is a polymer with good film forming and physical properties, easy to process. In the presence of nanostructured filler in a PVA matrix, the different polymer relaxation processes are affected, modifying the polymer mechanical response [23]. The UFM data on PVA / STO reported here improve our understanding of STO and PVA interactions, and hence provide a major advantage for the optimization of their technological applications. Experimental data on the nanoscale ultrasonic response in these materials illustrate the potential of UFM and contribute to the development of ultrasonic-AFM techniques.

**2- Materials and methods**

PVA in granular form (MW 31000-50000, 98-99% hydrolized), and STO NPs (of ≈ 100 nm in diameter) were purchased from Sigma Aldrich. For the preparation of PVA / STO nanocomposites, the PVA granules were dissolved in milli-Q water by consistent stirring at ≈ 60 ºC. STO NPs were very slowly added to the PVA solution, up to a total PVA + STO concentration of 6.25 w% in the solution. The heating temperature of the mixture was then lowered to 50 ºC, and kept stirring for 2 hours at this temperature. Eventually, the mixture was poured into a Petri dish, and kept at room temperature for 36 hours, appropriately covered to control evaporation. At this stage, a PVA / STO film of a few microns in thickness could be easily peeled out from the glass container. PVA / STO films in 30/70 and 70/30 w/w% were prepared. PVA films without NPs were also prepared in a similar way, lowering the temperature of the PVA solution to room temperature once the PVA granules were totally dissolved. Due to the addition of STO, the PVA samples, transparent in pure form (in the absence of nanoparticles), acquired an opaque white color. Shapes of ≈ 1x1 cm2 were cut from the films, mounted on an appropriate sample holder, and used for the AFM/UFM measurements.

To implement UFM, a standard commercial AFM (NANOTEC) was appropriately modified [2, 4]. Data acquisition and analysis were carried out using the WSxM software [24]. UFM was typically performed at ultrasonic frequencies of ≈ 3.8 MHz and modulation frequencies of 2.4 KHz. Olympus Silicon Nitride cantilevers with a pyramid-like tip shape and nominal spring constant of 0.06 Nm$^{-1}$ were used for the AFM/UFM experiments.

**3- Results and discussion**

Before initiating the discussion about the AFM/UFM results obtained on the PVA / STO hybrid films, data on PVA films prepared without STO nanoparticles (pure PVA) are briefly discussed for the sake of comparison. Fig. 1 (a) (b) shows topographic contact-mode AFM and UFM images simultaneously recorded on pure PVA over a same surface region. Fig. 1 (c) displays contour lines along the arrows indicated the images. The PVA surface is quite flat in the considered area, with height variations of ≈ 6 nm. In Fig. 1 (b), nanoscale variations of the UFM contrast are apparent. Scattered rounded areas of lower contrast with diameters < 150 nm are resolved in UFM, not straightforwardly correlated with the topography, which can be related to differences in the PVA local density. Previous AFM studies using Single Molecule Force Spectroscopy indicate that the elastic properties of PVA molecules scale linearly with their contour length [25]. Some slight scratches evident in both the AFM and UFM images in Fig. 1, might be due to the fact that when scanning in the contact mode some polymer chains adhere to the tip, particularly if this and/or the sample surface are not totally clean, and are pulled ahead as the tip moves. The structural arrangement of the PVA molecular chains within the films is expected to be dependent on the film preparation process.

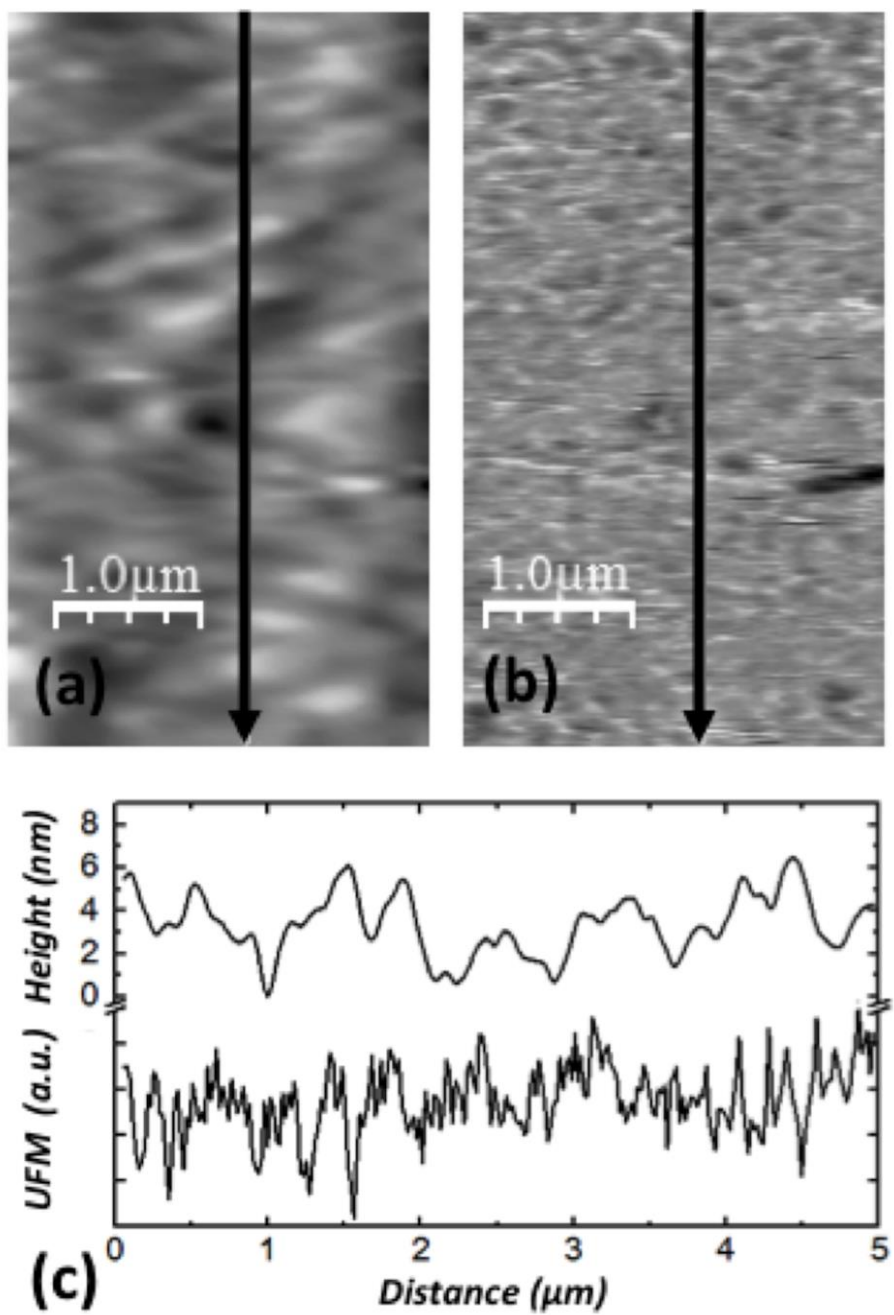

FIGURE 1. Pure PVA. (a) Topography in contact-mode AFM (b) UFM image simultaneously recorded with (a), over the same surface area. (c) Contour lines along the arrows in (a) and (b).

Fig. 2 (a) shows a tapping-mode AFM topograph of the STO NPs deposited on a silicon substrate. To prepare this sample, STO NPs were poured in distilled water, and consistently stirred. A droplet of the mixture was deposited onto a silicon surface, which was kept in vacuum until the liquid was evaporated. Fig. 2 (b) corresponds to a topographic contour along the arrow indicated in Fig. 2 (a). Height variations up to ≈ 130 nm are observed in the image. Clusters of ≈ 350 nm in diameter are clearly resolved in Fig. 2 (a). Taking into account that the nominal NP size is of ≈ 100 nm in diameter, the clusters are formed by ≈ 3-5 NPs. From both Fig. 2 (a) and (b), it is clear that the NPs gather in small clusters, which then join to form larger aggregates of ≈1.2 μm in diameter.

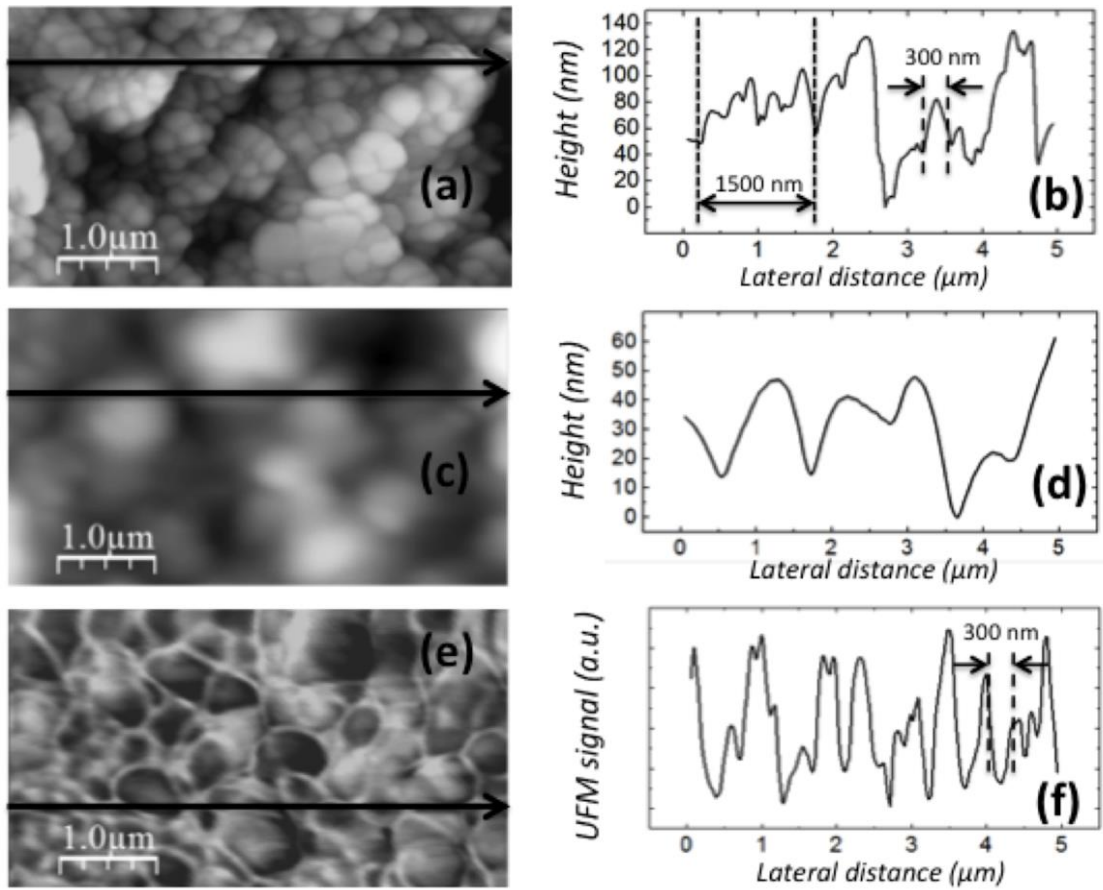

FIGURE 2. (a, b) STO NP deposited on a Silicon substrate. (a) Topography in tapping-mode AFM. (b) Contour along the arrow in (a). (c-f) PVA / STO composite in 30 / 70 w/w%. (c) Topography in contact-mode AFM. (d) Contour along the arrow in (c). (e) UFM image. (f) Contour along the arrow in (e).

Fig. 2 (c) is a contact –mode AFM topograph of a PVA / STO composite sample in 30/70 w/w%. Fig. 2(d) is a topographic contour along the arrow in Fig. 2 (c). In the composite, the surface roughness is lower than in Fig. 2 (a). Height variations are limited to ≈ 60 nm, and the small NP clusters are not distinguished in most cases, being only the larger NP aggregates well resolved. Nevertheless, the NP clusters are still well resolved in Fig. 2 (e), which corresponds to an UFM image recorded on the composite sample. Fig. 2 (f) is an UFM signal contour along the arrow in Fig. 2 (e).

According to the obtained results, we understand that the PVA polymer in the composite sample covers the NP clusters, filling the empty spaces among them. Typically, UFM provides extreme sensitivity to topographic changes, which lead to changes of the tip–sample contact area, and hence of the contact stiffness. Nevertheless, in our case, variations of the tip–sample contact area due to surface roughness (Fig. 2(c, d)) cannot account for all the observed UFM features (Fig. 2(e, f)). Some of the NPs in Fig. 2e may be located very near the surface, subjected to the tip-induced stress field as the tip periodically indents the sample surface at ultrasonic frequencies. Also, nanoscale ultrasound propagation might be closely dependent on the morphology of the buried nanostructures. The precise way in which the edges of the buried nanostructures may influence the UFM signal is still the subject of research.

A comparison of Fig. 2 (b), (d) and (f) indicates that the STO NPs aggregate in the same manner independently of whether they are embedded in the polymer matrix or lying on the silicon substrate. This suggests that during the preparation of the composite in the PVA solution the NP-PVA interactions are not sufficiently strong to substantially modify the NPs arrangement. When trying to record contact-mode

AFM/UFM images of the sample of STO NPs deposited on silicon, in the absence of the polymer, the NPs were swept away by the tip. Hence, it was not possible to perform UFM on this sample. For the composite sample, the UFM image (Fig. 2(e)) displayed a rich variety of contrast.

Fig. 3 (a) and (b) are topographic contact-mode AFM, and UFM images, respectively, recorded over a same surface region of a PVA / STO composite sample in 30 / 70 w/w%. Regions with characteristic UFM contrast have been enclosed with circles in both the topographic and the UFM images. In the region labelled A, some STO NP clusters distinguished in the topography (Fig. 3(a)) can be directly correlated with a specific UFM contrast (Fig. 3 (b)). However, clusters such as *i* and *ii* that are topographically similar, give rise to totally different elastic contrast, namely darker and brighter then their surroundings respectively. In UFM, a darker (brighter) contrast is typically indicative of a lower (higher) contact stiffness. Young's Modulus of STO is expected to be two orders of magnitude higher than that of PVA (Young's Modulus of STO ≈102 GPa, Young's Modulus of PVA ≈1 GPa) [26]. Nevertheless, apparently, we are measuring different rigidities for different STO NP clusters within the polymer. We attribute these results to the fact that the nanoscale response of the STO NP to the ultrasonic vibration will be highly dependent on the STO environment. Bonding and interactions at the PVA/STO interface in the presence of ultrasound may lead prevent the propagation of ultrasound and lead to a lower UFM contrast in the images. For the current STO concentration (70% wt.), STO NP aggregates might be linked across the entire thickness of the composite film. Individual STO clusters could be found either in close contact (or separated by a very thin PVA layer at most) or relatively isolated, surrounded by the PVA matrix (see Fig. 3(c)). In this latter case, a lower UFM NP contrast is expected, assuming that ultrasound is damped or scattered at PVA/STO interfaces.

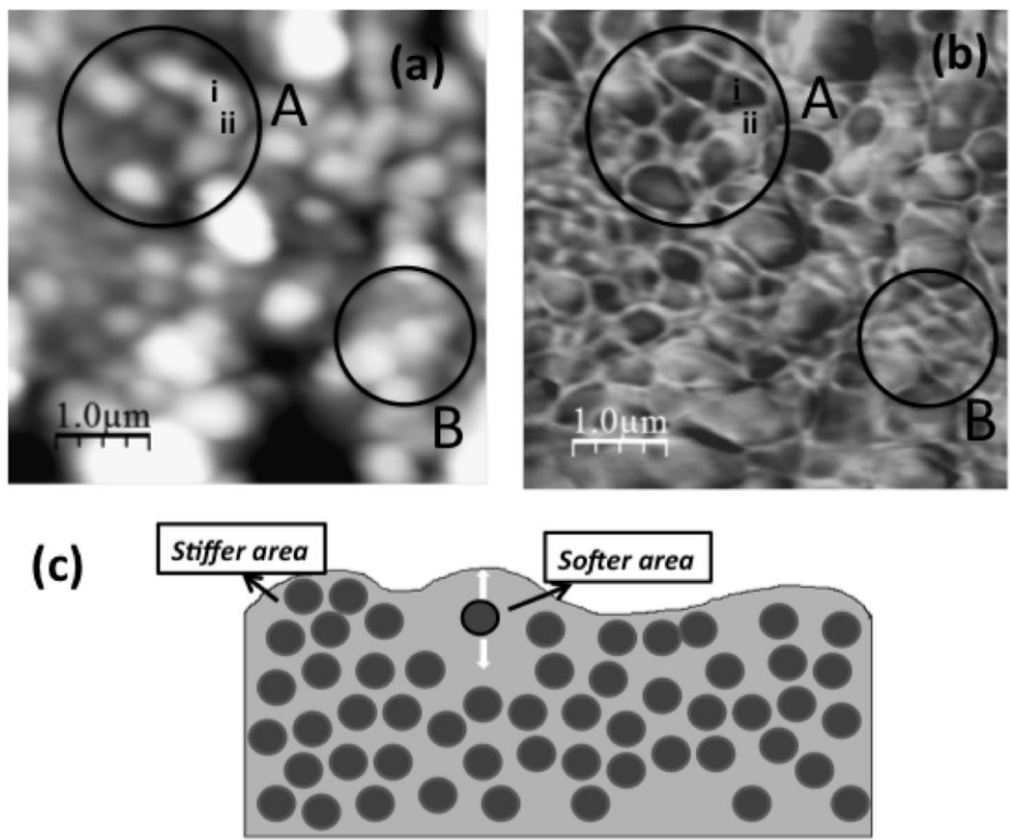

FIGURE 3. PVA / STO composite in 30 / 70 w/w%. (a) Topography in contact-mode AFM. (b) UFM image recorded simultaneously with (a), over the same surface area. The circles enclose regions with characteristic features, discussed in the text. (c) Schematics of possible distributions of STO nanoclusters inside a PVA matrix.

In the region labelled B, in the topography, we also distinguish topographic protrusions similar to those in the region labelled A. Nevertheless, here the UFM contrast is not straightforwardly correlated with the protrusions. Inside areas corresponding to single topographic protrusions, differences in elastic contrast are apparent. Again, we assign the origin of such contrast to the PVA / STO interface properties that alter the propagation of ultrasound on the nanoscale, and hence give rise to different UFM responses. In addition, inhomogeneities in the thickness or density of the PVA layer surrounding the clusters might also contribute to the image contrast. Fig. 3 (c) sketches a variety of nanocluster distributions inside a polymer matrix that illustrate distinct cases that may be accounted for to explain the rich variety in UFM responses in the composite sample images.

Fig. 4 shows topographic contact-mode AFM (Fig. 4 (a)), UFM (Fig. 4(b)), and Friction Force microscopy (FFM) images in forward (Fig. 4(c)) and backward (Fig. 4(d)) scans recorded over a same surface area of a PVA / STO composite sample in 30 / 70 w/w%. Fig. 4 (a) and Fig 4 (b) were simultaneous recorded, and Fig. 4 (c) and Fig. 4 (d) were recorded immediately after, in the absence of ultrasound. Regions with characteristic UFM contrast have been outlined with circles in all the images to facilitate the comparison of the different responses. The relationship between the topography and the UFM contrast here resembles this in region B from Fig. 3 (a) (b). From Fig. 4 (c) and (d) it is apparent that no clear frictional contrast can be appreciated in these regions. The absence of a significant frictional response indicates that PVA is completely covering the STO NP clusters in the composite, and confirms that the UFM contrast in these samples stems from buried locations.

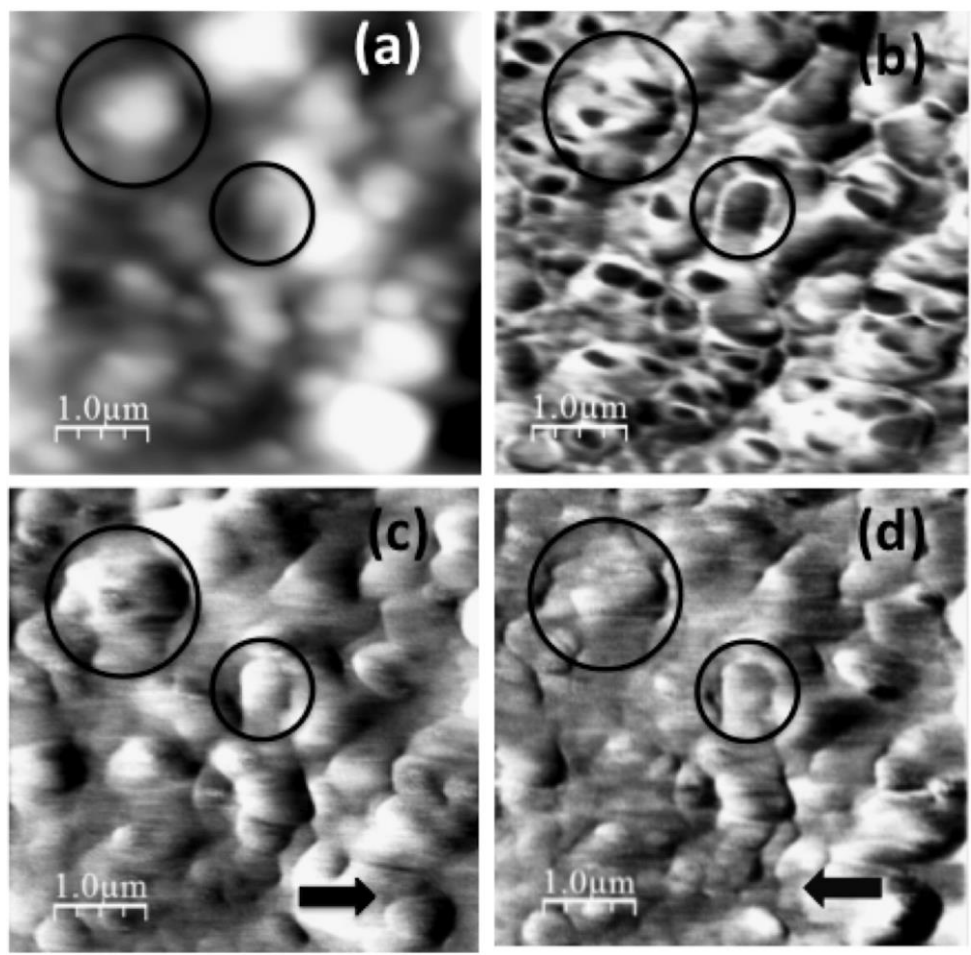

FIGURE 4. PVA / STO composite in 30 / 70 w/w%. (a) Topography in contact-mode AFM. (b) UFM image simultaneously recorded with (a), over a same surface area. (c, d) FFM images recorded in forward (c) and backward (d) scans immediately after recording (a) and (b), over the same surface area. The circles enclose regions with characteristic features, discussed in the text.

Fig. 5 corresponds to contact-mode AFM (Fig. 5(a)) and UFM (Fig. 5(b)) images simultaneously recorded over a same surface area on a PVA / STO sample in 70 / 30 w/w%. Height variation in the topographic image is now limited to ≈ 24 nm. The surface roughness is much lower than on the composite samples with a higher STO concentration. It is apparent from these images that the lower STO concentration prevents the formation of the large NP aggregates observed in Fig. 2. The size of the topographic protrusions in Fig. 5 (a) ranges between 100-200 nm in diameter, being 100 nm in diameter the nominal NP size. The UFM image in Fig. 5 (b) provides a characteristic contrast for the NP protrusions, allowing us to identify the NPs easier than in the topography. According to the size of the protrusions, they must correspond to NP clusters formed by just 2 or 3 NPs, or even to individual NPs, being smaller on average then those formed in the PVA / STO sample in 30 / 70 w/w% (Fig. 2). The absence of large NPs aggregates, and the lower NP cluster size may be understood as due to the fact that the higher PVA concentration in the solution during the formation of the composite hinders the NPs mobility when the temperature is diminished. The lower (darker) UFM contrast in Fig. 5 (b) at the areas where the topographic protrusions in Fig. 5 (a) are located indicates the presence of subsurface STO NPs, surrounded by the PVA matrix. Notice that not all the NPs or NP clusters yield the same type of UFM contrast in the image. For instance, cluster labeled *i* appears with a much darker contrast than the one labeled *ii*, on which the contrast is only slightly darker then on its surroundings. Notice that in region labeled *iii*, the (dark) UFM contrast indicates the presence of a buried cluster that is not correlated to a topographic protrusion. As remarked above, the fact that the STO NPs yield a darker (lower) contrast in UFM, indicative of apparently softer regions is, in principle, not expected. So far STO is stiffer than PVA, one would anticipate that the rigidity (UFM signal) should be higher (brighter) over the NPs. The contrast mechanism in subsurface AFM is dependent both on the acoustic wave propagation through the sample, and on the tip-sample interactions. When performing UFM, an effective acoustic field establishes inside the sample, and the sample atoms are subjected to mechanical vibration accordingly. Rayleigh scattering of the acoustic wave from individual NPs has been suggested to explain some subsurface AFM experiments reported in the literature [27]. Our interpretation of the current data is based on the hypothesis that the ultrasonic vibration of PVA-STO interface atoms can be strongly damped. This causes a disruption of the acoustic wave field at the NP locations, and hence the UFM signal diminishes. We verified experimentally that input of the ultrasonic excitation from the back of the sample, or from the cantilever base, when implementing UFM [28] did not result in any significant qualitative differences in the UFM NPs image contrast on regions similar to this in Fig 5 (a), (b). NPs located near to the surface will be subjected to the tip-induced stress field, but still the PVA-NP interface properties may reduce the expected UFM contrast. In fig. 5 (a) (b) those NPs that yield a lower UFM contrast in (b) cannot be appreciated as topographic protrusions in (a) in most cases (see e.g. particles type i and iii), which indicates that they should be located deeper, more distant from the sample surface, being presumably less affected by the tip-induced stress field.

Images with higher magnification in Fig. 5(c) and (d) were simultaneously recorded over the same surface area on a PVA/STO composite sample in 70/30 w/w%. Fig. 5 (c) is the derivative of a contact-mode AFM topographic image, and Fig. 5 (d) an UFM image. In Fig. 5 (c) the derivative is displayed instead of the topography to permit a better appreciation of the gradient variations. It is clear from Fig. 5 (d) that the topographic protrusions yield darker contrast that their surroundings in UFM. Furthermore, here it can clearly be appreciated that those protrusions in UFM appear delimited by a thin halo of darker UFM contrast, of nonuniform thickness. This kind of halo can be also found in most of the NP clusters in Fig. 6 (b). We understand that this lower-contrast halo indicates a hindrance to ultrasound propagation at the PVA/STO interface. Damping of the ultrasonic signal at the PVA/STO interface provides a plausible explanation for our data. We are currently investigating possible damping mechanisms, which might be related to dissipative motions of the PVA chains at the STO interface, perhaps dependent on the presence or absence of water at the PVA/STO interface region or on the piezoelectric character of the STO NP.

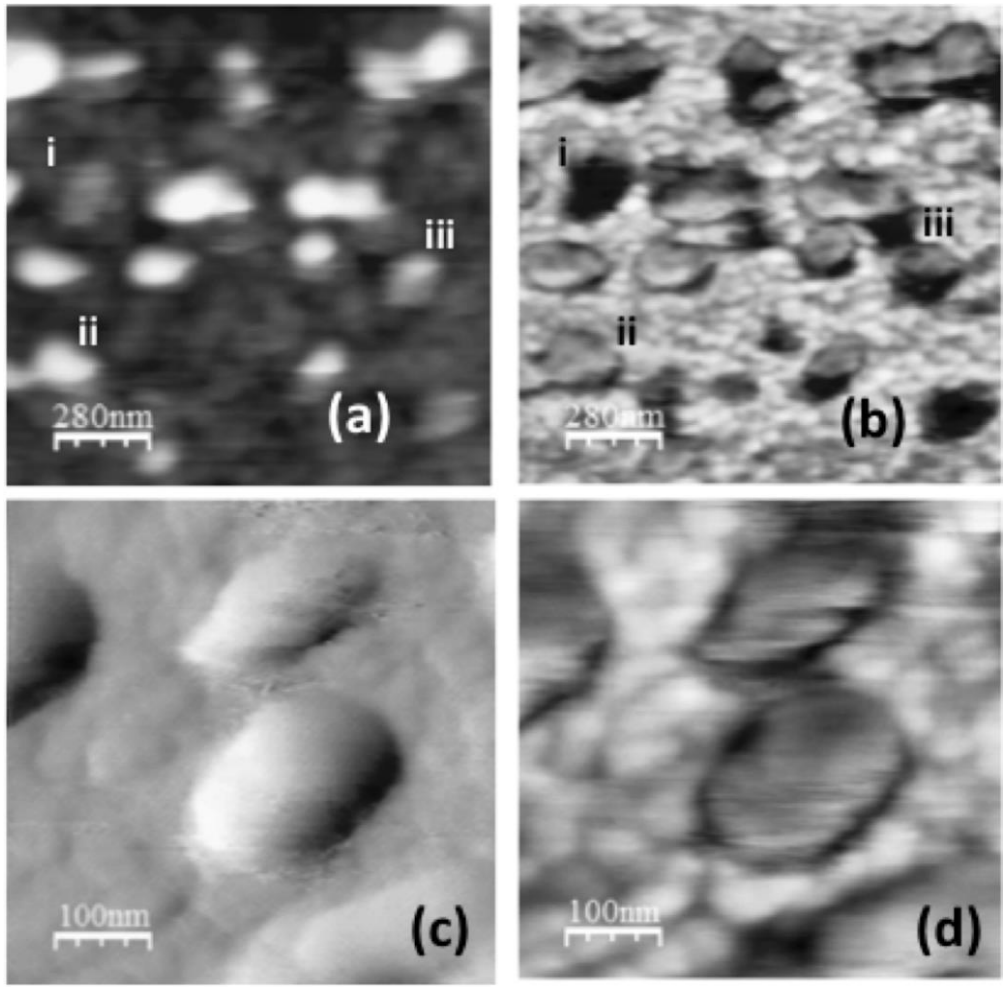

FIGURE 5. PVA / STO composite in 70 / 30 w/w%. (a) Topography in contact-mode AFM. (b) UFM image recorded simultaneously with (a), over a same surface area. (c) Topography (derivative image) in contact-mode AFM (d) UFM image recorded simultaneously with (c), over a same surface area.

Fig. 6 provides further evidence for our conclusions. Fig. 6 (a) and (b) correspond to a PVA / STO composite sample in 30 / 70 w/w%; (a) is an UFM image, and (b) is the UFM signal contour along the arrow in (a). Fig. 6 (c) and (d) correspond to a PVA / STO composite sample in 70 / 30 w/w%; (c) is an UFM image, and (d) is the UFM signal contour along the arrow in (c). Here, in Fig. 6(a) and (c), the UFM contrast allows us to distinguish the STO clusters embedded in the PVA matrix. In the case of the sample with the lower STO NP concentration (Fig. 6(c)), individual NPs or NP couples are resolved, but their tendency to aggregate together is noticeable.

The dashed lines at the arrows in the images indicate a same lateral extension, which correspond to a single NP cluster in Fig. 6 (a, b), and to several smaller clusters in Fig. 6 (c, d). In Fig. 6 (a) and (c), the NPs clusters appear with the aforementioned characteristic darker halo in the UFM contrast. This is also apparent from the signal contours in Fig. 6 (b) and (c). We understand these halos as an evidence that the PVA / STO interfaces may effectively damp ultrasonic vibration, thereby modifying the expected stiffness-related UFM NP response.

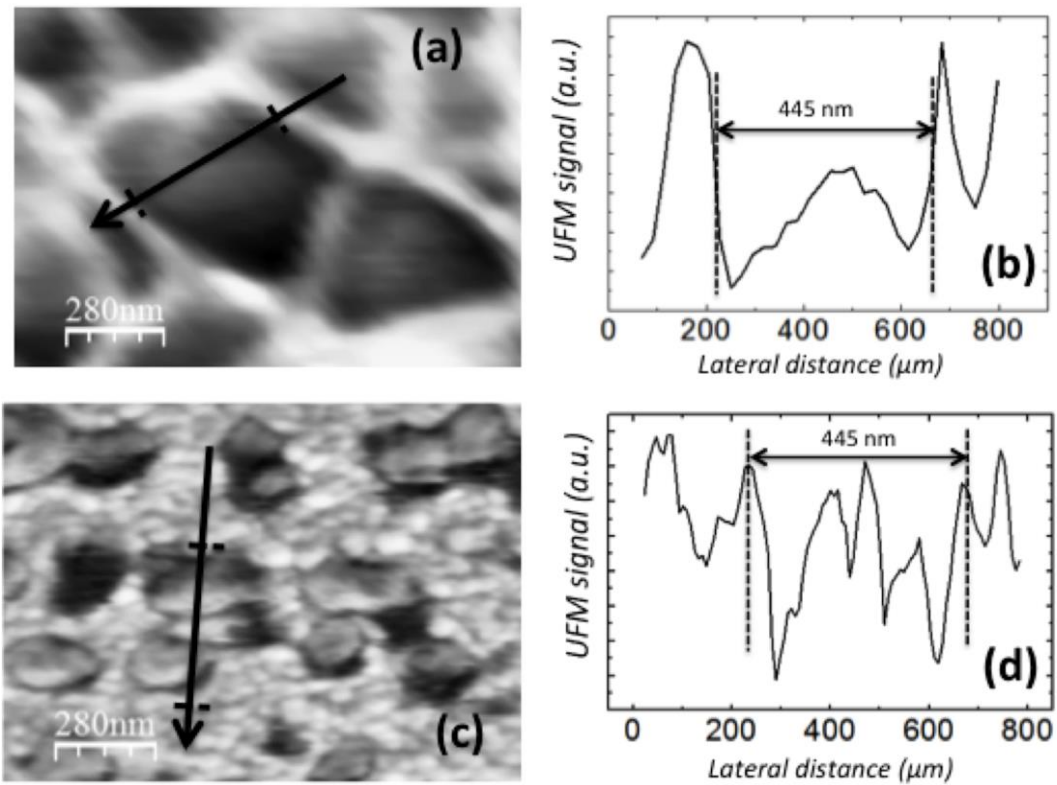

FIGURE 6. (a, b) PVA / STO composite in 30 / 70 w/w%. (a) UFM image. (b) UFM signal contour along the arrrow in (a). (c, d) PVA / STO composite in 70 / 30 w/w%. (c) UFM image. (d) UFM signal contour along the arrow in (c).

**4- Conclusions**

Summarizing, we have applied UFM to characterize PVA / STO nanocomposite samples. The STO NPs are better resolved using UFM than using contact-mode AFM. The individual STO NPs form nanoclusters, that gather in nanoaggregates for high STO concentrations. Similar aggregation is observed for NPs in the composite samples than for NPs deposited on top of a flat silicon substrate from a milliQ water solution in the absence of polymer. The absence of frictional contrast in AFM for STO aggregates, well-resolved in UFM, allows us to conclude that those are covered by PVA. The PVA/STO samples gave rise to a rich variety of contrast in UFM. Most STO NPs in the composite samples exhibited a lower (softer) UFM contrast than the PVA matrix. In addition, a lower contrast halo could be resolved at those NPs with intermediate UFM contrast. The data are interpreted by considering that ultrasound can be strongly damped at PVA/STO interfaces on the nanometer scale. Our experimental results demonstrate that the nanoscale ultrasonic response at the PVA/SrTiO3 interface plays a fundamental role in the UFM image contrast, and emphasize the UFM capability to resolve ultrasound-induced features on nanocomposite samples with nanoscale resolution.

ACKNOWLEDGMENT. We thank N. Kumar for assistance in some measurements in the initial stage of this study. S. M. acknowledges financial support (postdoctoral grant) from the Junta de Comunidades de Castilla-La Mancha (JCCM), and G. M. J. and A. L. acknowledge financial support from the University of Castilla-La Mancha (UCLM) for stays in the Laboratory of Nanotechnology in Almaden (Spain). Financial support from the Junta de Comunidades de Castilla-La Mancha (JCCM) under project PBI-08-092, and the Spanish Ministry of Science and Innovation (MICINN) under project MAT2009-08534 is gratefully acknowledged.